\documentstyle[12pt,epsf]{article}
\def\beq{\begin{equation}} 
\def\eeq{\end{equation}} 
\def\bea{\begin{eqnarray}} 
\def\eea{\end{eqnarray}} 
\def\nn{\nonumber} 

\begin{document}

\newcommand{\raltitle}
{Atmospheric neutrino oscillations and maximal $\nu_{\mu}-\nu_{\tau}$ mixing in
unified models}

\newcommand{\ralauthor}
{B. C. Allanach$^1_a$}

\newcommand{\raladdress}
{$^1$ DAMTP, Silver Street, Cambridge, Cambs., CB3 9EW, U. K.}

\newcommand{\ralabstract}
{The recent data describing the evidence for neutrino oscillations by Super
Kamiokande implies that two neutrinos are very strongly mixed.
We ask the question: can approximate maximal neutrino mixing be natural in
unified (GUT or string) models? We attempt to answer this question in as much
generality as possible. Without specifying a particular model, we are able to
show that a gauged family symmetry can naturally provide the required maximal
mixing between $\mu$ and $\tau$ neutrinos. The second and third family
neutrino mass eigenstates are almost degenerate with masses of order 0.2 eV.}

\begin{titlepage}
\begin{flushright}
hep-ph/9806294\\
\end{flushright}
\vspace{.2in}
\begin{center}
{\large{\bf \raltitle}}
\bigskip \\ \ralauthor \\ \mbox{} \\ {\it \raladdress} \\ \vspace{.5in}
{\bf Abstract} \bigskip 
\end{center} 
\setcounter{page}{0}
\ralabstract
\vfill \noindent
{\tt a) b.c.allanach@damtp.cam.ac.uk}
\end{titlepage}

\section{Introduction}
The recent data from the Super-Kamiokande
collaboration~\cite{Fukuda:1998ub,CKJ} provides
compelling evidence 
for the existence of neutrino
oscillations~\cite{CKJ} and therefore also neutrino masses. We take the
simple view here that  to a good
approximation the
data are the result of only two neutrino flavours oscillating. This is 
true if mixing angles between any other neutrino species are small.
There are strong experimental hints that this is indeed the case:
the LSND
results~\cite{Athanassopoulos:1997pv}, and the
small-angle MSW solution~\cite{MSW} to the solar neutrino problem.
$\nu_e-\nu_\mu$ oscillations as an explanation of the atmospheric neutrino
effect are disfavoured by the CHOOZ
data~\cite{Gonzalez-Garcia:1998vk,Apollonio:1997xe,Giunti:1998qn}
and by the ratio of upward to downward events measured by
Super-Kamiokande for $\mu$-like and $e$-like signatures~\cite{Fukuda:1998ub}.
The atmospheric anomaly could then be due to $\nu_\mu$ oscillations involving
a sterile
neutrino $\nu_s$ and/or $\nu_\tau$~\cite{Lipari:1998rf,Foot:1998iw}.
In the following, we assume that $\nu_\mu$ mixes dominantly with $\nu_\tau$
rather than $\nu_s$. This hypothesis could in principle be
checked by examining the ratio of charged to neutral current
events~\cite{Vissani:1997pz} or the up-down ratio of contained inclusive
multi-ring events~\cite{Hall:1998dx}. 

To interpret the atmospheric data as an oscillation between two neutrinos, we
require $\Delta m^2 \sim 5-50 \times$ 10$^{-4}$ eV$^2$ and
$\sin^2 2 \theta \sim 1$~\cite{Gonzalez-Garcia:1998vk}. 
Naively, one might think there is a conflict between the usual predictions of
unified theories and these parameters. 
Quark-lepton unification at some high
(GUT or string) scale is a common prediction of unified theories.
This would imply that the mixings of quarks and their leptons be
equal to within factors of order 3 or so, which could arise from
renormalisation effects.
This is of course in conflict with charged fermion mass/mixing data and is so
too simplistic. Non-renormalisable operators involving unified Higgs can be
employed to explain 
further factors of around 3, leading to successful predictions of
both
GUT~\cite{GJ} and string-inspired~\cite{Allanach:1997hz} models. Even within
this context, it is 
necessary to understand where the mass suppression of the neutrinos with
respect 
to the charged fermions arises, 
because the Dirac neutrino masses are 
equal to the up-quark masses at the unification scale. 
This suppression has a natural explanation in
terms of
the see-saw mechanism. 
In the see-saw mechanism,
heavy right-handed neutrinos of mass much larger than the electroweak scale
are introduced which suppress the effective light neutrino masses.
These right handed
neutrinos are naturally present as
the partner of the right handed electron
in models containing the symmetry SU(2)$_R$, such as left-right symmetric
models~\cite{sk}, GUTs~\cite{Albright:1998vf} and
string-inspired
models~\cite{us}. Their mass terms are not constrained to be the same
magnitude as that of the quarks and so their large size (as well as hierarchies
between them) can be well motivated~\cite{us}. In fact, their size is often
related to the scale of SU(2)$_R$  breaking (or the breaking of a group of
which SU(2)$_R$ is a subgroup).

Another problem is to generate such a large (maximal) mixing angle
when the quarks are known to have small mixings, as described by the CKM
matrix. There have been some specific proposals to solve this problem,
for example by imposing discrete family
symmetries~\cite{Xing:1998th,Drees:1998id}.
In ref.~\cite{Bando:1998ns}, a phenomenological texture is assumed for the
Dirac and Majorana mass matrices of the $(e,\mu,\tau)$ neutrinos which
reproduces large mixing between $\nu_\mu$ and $\nu_\tau$. 
The authors use the Georgi-Jarlskog ansatz~\cite{GJ} for down-quark and lepton
unification, which naively produces a $\sin^2 2 \theta_{e \mu}$ mixing angle
too big to satisfy the small angle MSW explanation of the solar neutrino
problem. 
However, if in this scheme the $\nu_\mu, \nu_\tau$
neutrinos are maximally mixed, the $\sin^2 2 \theta_{e
\mu}$ mixing angle is brought into line with what is required.
The form of the
down-quark Dirac mass matrix is different to that of the up-quark Dirac matrix
and so would not seem to be a natural prediction of theories that unify up and
down quarks, as is often the case.
In ref.~\cite{Binetruy:1996cs}, a large mixing is generated by using a
spontaneously broken gauged U(1)$_F$ family symmetry~\cite{gaugesym}. 
This works very well and in refs.~\cite{Binetruy:1996cs,neutpeeps}, examples
that generate large mixings were found. This approach will be utilised later
to motivate a class of texture.
The implications of this family
symmetry for the neutrino mass structure was also investigated in
ref.~\cite{Dreiner:1995ra}.
Another recent specific SO(10)
model~\cite{Albright:1998vf} has maximal mixing of the charged $\mu$ and
$\tau$ leptons.

Our basic assumption is that the
Dirac lepton/neutrino mass matrices have
small mixings, comparable to that of the quarks. This is motivated by
quark-lepton unification.
It is then natural to propose that the observed maximal mixing is due to
maximal mixing of the heavy right handed neutrinos.
Because the mixing in the Majorana sector is not related
by the vertical gauge symmetry to that of the quarks, it can have a different
form. This then filters through the
see-saw mechanism to provide maximally mixed light left-handed neutrinos.
Here, we analyse a specific class of texture in
detail. 
We check that the corrections expected from our approximations
do not change the qualitative result.
We then provide physical motivation for this ansatz in
terms of a U(1)$_F$ gauged family symmetry {\em a la}\/
refs.~\cite{Binetruy:1996cs,neutpeeps}, and derive constraints on
the 
quantum numbers of the right-handed $\nu_\tau, \nu_\mu$.

\section{Majorana Mass Ansatz}

Here, we postulate that mass terms involving $\nu_{e,s}$ don't affect the
atmospheric neutrino oscillations, allowing us to concentrate on $\nu_\mu,
\nu_\tau$ masses only.
The outcome of the see-saw mechanism is two light approximately left-handed
neutrinos whose Majorana mass matrix is approximately
\beq
m_\nu = - m_D^T M^{-1}_R m_D, \label{usual}
\eeq
where $m_D$ is the Dirac neutrino mass matrix of $\nu_\mu$ and $\nu_\tau$ (set
equal to the charm-top quark mass 
matrix through quark-lepton unification) and $M_R$ is the Majorana mass matrix
of the right-handed $\nu_\mu, \nu_\tau$. 
We expect that the charm-top
quark mass matrix is approximately diagonal in the SU(2)$_L$ current basis.
Otherwise, the
measured smallness of $|V_{cb}| \sim 0.03$ would require a large
cancellation with an element in the down quark-sector (which we consider
unnatural\footnote{However, this possibility could be motivated by a suitable
weakly broken discrete symmetry~\cite{demo} but we do not consider this
case here.}). Thus $m_D$ is of the order
\beq
m_D \sim \left( \begin{array}{cc} m_c & 0 \\ \xi m_t & m_t \\
\end{array} \right), \label{mango}
\eeq
where $\xi \sim O(|V_{cb}| / 2) \ll 1$.
The zero in the (1,2) position of $m_D$ is only approximate: a non-zero
element has negligible effect if it is much smaller than $m_t$. $m_D$ is
approximately diagonal in the sense that the orthogonal diagonalising
matrices $U,V$ defined by
\beq
U^T m_D V = \mbox{diag} (m_c, m_t)
\eeq
are approximately $1_2$.
In many models, see for example ref.~\cite{us}, Eq.(\ref{mango}) is a
prediction valid at 
a particular high scale rather than an order of magnitude statement.
$M_R$ is the two by two right-handed
neutrino Majorana mass matrix whose eigenvalues are expected to be much
greater than $m_t$. 

We now postulate that the form
\beq
M_R \approx \left( \begin{array}{cc} 0 & M \\ M & 0 \\ \end{array} \right) 
\label{rhmaj}
\eeq
will provide maximal mixing between the light $\nu_\tau, \nu_\mu$.
This possibility has already been included in some of the models of
ref.~\cite{Binetruy:1996cs,neutpeeps}.
Using Eq.(\ref{usual}), we determine the effective light neutrino mass matrix
\beq
m_\nu \approx m' \left( \begin{array}{cc} 2 \xi   &
1 \\ 1  & 0 \\ \end{array} \right),  \label{lhmix}
\eeq
where $m' \equiv m_c m_t / M$.
We can already see the property of maximal mixing from Eq.~\ref{lhmix} from
the smallness of diagonal entries in comparison to the off-diagonal ones.
We now have two approximately degenerate neutrinos, with masses
\bea
m_2 &=& m' (1 - \xi + O(\xi^2)), \nn \\ 
m_3 &=& m' (1 + \xi + O(\xi^2))
\eea
respectively. 
Therefore, $\Delta m_{23}^2 \approx 4{m'}^2 \xi + O(\xi^3)$. Using $\Delta
m_{23}^2 = 3 \times 10^{-3}$~eV$^2$ to explain the atmospheric data and
substituting $\xi =|V_{cb}|/2$ yields 
$m' \sim O(0.2)$~eV. This value of  $m'$ corresponds to $M \sim
O(10^{12})$~GeV.
The energy
density of relic neutrinos is~\cite{Sarkar:1997ki}
\beq
\Omega_\nu h^2 = \sum_i \frac{m_{\nu_i}}{94 \mbox{~eV}} \times 4.3 \mbox{~}
10^{-3}
\eeq
where h = H/100 Km/sec/Mpc is the present Hubble parameter. 
So for $h > 0.5$, one has $\Omega_\nu < 0.017$, i.e.\ there
would be at most a 2\% component of hot dark matter if $\Omega_{total} = 1$.

We may now ask  how far from maximal mixing the neutrinos are.
$m_\nu$ is diagonalised by the 2 by 2 orthoganol rotation
$U_\nu^T m_\nu U_\nu$, where
\beq
U_\nu = \frac{1}{\sqrt{2}}
\left( \begin{array}{cc} 1+\xi/2  & -1 + \xi/2 \\ 1 -
\xi/2  & 1 +
\xi/2 
\\ \end{array} \right) + O(\xi^2).
\eeq
The full lepton mixing comes from $U_\nu^T U_L$, where $U_L$ is the
analogous
mixing of the $\mu$ and $\tau$ 
leptons, which is supposed to also be small as a requirement of quark-lepton
unification. The off-diagonal elements of $U_L$ are of order $\xi$ and so
we parameterise $U_L$ by
\beq
U_L = \frac{1}{\sqrt{1 - a \xi^2}} \left( \begin{array}{cc} 1 & a\xi \\ - a\xi
& 1 \\ \end{array}
\right)
\eeq
where $a \sim O(1)$. This predicts that 
\beq
\sin^2 2\theta_{\mu \tau} = 1 
+ (1/\sqrt{2}-1/2+2a-a^2) \xi^2 + O(\xi^3),
\eeq 
thus the mixing is maximal
up to effects of order $|V_{cb}|^2$.

We now consider corrections to texture zeroes in $M_R$.
If we perturb the zeroes:
\beq
M_R = M \left( \begin{array}{cc}
c & 1 \\ 1 & b \end{array} \right),
\eeq
we find that 
\beq
m_\nu \approx 
\frac{1}{1-cb} m'
\left( \begin{array}{cc}
2 \xi - b \frac{m_c}{m_t} -c \xi^2 m_t/m_c & 1 - c\frac{m_t}{m_c}\xi \\ 
1 - c\frac{m_t}{m_c}\xi & -c \frac{m_t}{m_c} \end{array}
\right). \label{correff}
\eeq
$\xi m_t / m_c \sim O(1)$, so we see that maximal mixing is spoiled by the
(2,2)
element of Eq.(\ref{correff}) unless $c < O(\xi)$.
In this case, we obtain the same form of $m_\nu$ as Eq.(\ref{lhmix}) for
$b<O(1)$.
One should also check that the form of $m_\nu$ in Eq.(\ref{lhmix}) is not
spoiled by renormalisation (i.e.\ that large positive corrections don't appear
on the diagonal). It is necessary to set the model valid at scales below 
SU(2)$_R$ breaking 
to make this check, so we merely examine a few common examples here.
If the unified theory breaks straight to The Standard Model, radiative
corrections are of the form~\cite{Chankowski:1993tx}
\bea
16 \pi^2 \frac{d m_\nu}{d \ln \mu} &=& \left( \frac{1}{2} \lambda_2 - 3 g_2^2 + 
\mbox{Tr}(6 Y_U^\dag Y_U + 6 Y_D^\dag Y_D + 2 Y_L^\dag Y_L) \right) m_\nu +
\nn \\ 
&& \frac{1}{2} (m_\nu Y_L^* Y_L^T) 
+ \frac{1}{2} (Y_L Y_L^\dag m_\nu), \label{SM}
\eea
where $Y_{U,L,D}$ are the 3 by 3 up-quark, charged lepton and down quark
Yukawa matrices respectively, 
$\lambda_2$ is the Higgs self-coupling and $\mu$ is the $\overline{MS}$
renormalisation scale.
The dominant term on the RHS of Eq.(\ref{SM}) is the one proportional to
$m_\nu$ because the top-Yukawa coupling, $(Y_U)_{33}$ is larger than all other
couplings. Thus we see that the dominant correction to an entry of $m_\nu$
is of the same order (and form), implying that the texture is not spoiled by
large corrections to diagonal elements from the dominant corrections. 
The 2 Higgs doublet model and MSSM have a similar
form~\cite{Chankowski:1993tx} (but with different coefficients) to
Eq.(\ref{SM}). Thus in these models, the above reasoning still applies if
$\tan \beta$ is not extremely high. If $\tan \beta$ {\em is}\/ high, say around
50 (as one would expect if up and down quarks are unified), one must examine
the last two terms of Eq.(\ref{SM}) because $h_\tau\equiv (Y_L)_{33} \sim
O(1)$. 
$Y_L$ should have a similar form to $Y_U$, which we parameterise as:
\beq
Y_L = h_\tau \left( \begin{array}{cc}
d \xi & 0 \\
e \xi & 1 \\ \end{array} \right)
\eeq
where $d,e \sim O(1)$.
Then the last two terms in $16 \pi^2 d m_\nu / d \ln \mu$ are of order
\beq
m' h_\tau^2 \left( \begin{array}{cc}
d^2 \xi^2 & \frac{1}{2} \\
\frac{1}{2} & de \xi^2 \\ \end{array} \right),
\eeq
so the off-diagonal elements could possibly reduce the degeneracy of the two
neutrinos, but this would have to be checked in detail for any specific model.

\section{Family gauge symmetry}

In the above section we showed how the existence of texture zeroes in the
right-handed neutrino Majorana mass matrix and quark-lepton unification
predicted approximately maximal 
mixing of $\nu_\mu$, $\nu_\tau$ provided $M \sim 10^{12}$~GeV. 
To describe the origin of these features, we now appeal to the existence of a
U(1)$_F$ family gauge symmetry, which has been shown to lead to the correct
hierarchies in the masses of charged fermions~\cite{gaugesym} and which are
common in string theories. The family symmetry is broken at a high scale by
Standard Model singlet Higgs vacuum expectation values (vevs) $\langle\theta
\rangle
\sim \langle \bar{\theta} \rangle$, where $\theta, \bar{\theta}$ have family
charges -1, +1 respectively. 
Hierarchies in the effective mass terms are realised by U(1)$_F$ invariant
non-renormalisable operators
$\bar{f}_L f_R H \epsilon^n$, where $f_L, f_R$ are generic left and right
handed Standard Model fermions and $H$ is the relevant Higgs field.
$\epsilon \equiv \langle \theta \rangle / M'$ is a parameter of order 0.1,
suppressed by the heavy mass scale $M'$ which could be the string
scale~\cite{us}, or some other mass scale of heavy particles in the field
theory~\cite{fixedmass}. $n$ is a non-negative integer calculated from the
powers of $\theta$ or
$\bar{\theta}$ that are required to make the operator invariant under
U(1)$_F$. The fermions are assigned family dependent charges such that the
observed charged fermion mass/mixing hierarchies are well reproduced.
This assignment can lead to model dependence: for example, one has to
specify the number of electroweak Higgs doublets and their charges.
There are many different assignments for the charged fermions which can
reproduce the correct hierarchies and we assume here that this is the case.
Thus we do not consider the
U(1)$_F$ quantum numbers of the charged fermions, just $q_\mu, q_\tau$ (those
of the right-handed $\nu_\mu, \nu_\tau$ particles respectively)
suffice. Because we don't consider these other quantum numbers, we also don't
consider the possible U(1)$_F$ anomalies.
We derive constraints on $q_\mu, q_\tau$ by the requirement that they
reproduce the texture zeroes of Eq.(\ref{rhmaj}).

\begin{figure}
\begin{center}
\leavevmode   
\hbox{\epsfxsize=2.5in
\epsfysize=2.5in
\epsffile{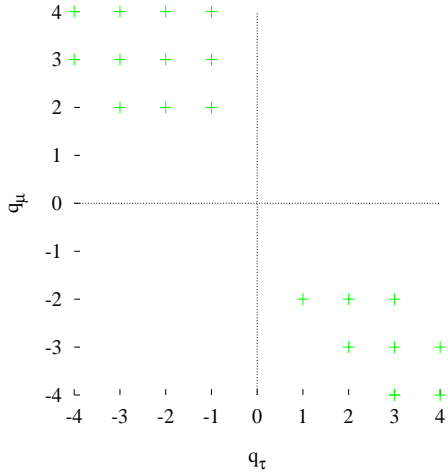}}
\end{center}
\caption{Possible integer charges of right handed neutrinos leading to
maximal mixing. A cross shows a charge assignment that leads to maximal mixing
of $\nu_\mu$, $\nu_\tau$.
We have only considered $|q_\mu|, |q_\tau| \le 4$.}
\label{Fig1}
\end{figure}

Thus we have the right handed neutrino masses of order
\beq
M_R \sim M''  \left( \begin{array}{cc} \epsilon^{2 |q_\mu|} &
\epsilon^{ |q_\mu+q_\tau|}  \\
\epsilon^{ |q_\tau+q_\mu|}  & \epsilon^{2|q_\tau|} \\
\end{array} \right), \label{noweresurfin}
\eeq
where $M''$ is another mass that sets the scale of
the right-handed neutrino masses (such as the GUT, string, or SU(2)$_R$
breaking scale).
Then, the requirement of approximate texture zeroes along the diagonal leads to
the simultaneous constraints
\beq
|q_\mu| > |q_\mu + q_\tau|/2 ,\ |q_\tau| > |q_\mu + q_\tau|/2. \label{const}
\eeq
These inequalities are incompatible if $q_\mu$ and $q_\tau$ have the same
sign, and so the
signs must be opposite. 
First, we examine the case where $q_\mu <0$ and $q_\tau>0$.
There are two relevant cases:
\bea
|q_\mu|\le q_\tau: &&  M_R=M'' \epsilon^{q_\tau-|q_\mu|} \left(\begin{array}{cc}
\epsilon^{3|q_\mu|-q_\tau} & 1 \\
1 & \epsilon^{q_\tau+|q_\mu|} \\ \end{array} \right), \nn \\
|q_\mu|>q_\tau: && M_R=M'' \epsilon^{|q_\mu| - q_\tau} \left(\begin{array}{cc}
\epsilon^{q_\tau+|q_\mu|} & 1 \\
1 & \epsilon^{3|q_\mu|-q_\tau} \\ \end{array} \right). \label{bottomline}
\eea
Requiring that the diagonal entries be smaller than the off-diagonal ones
now leads to the constraint
\beq
\frac{q_\tau}{3} < |q_\mu| < 3 q_\tau. \label{chargeoftheheavybrigade}
\eeq
For the other case $q_\mu>0$ and $q_\tau<0$, a similar analysis shows
that $\mu \leftrightarrow \tau$ in Eq.(\ref{chargeoftheheavybrigade}).
Whereas $b<O(1)$ is already satisfied, to reproduce $c<O(\xi)$ in
Eq.(\ref{correff}),
the (1,1) element must be at most of order $\epsilon^3$ in
Eq.(\ref{bottomline}). 
This further restricts possible charge assignments, the
result of which is shown in Fig.(\ref{Fig1}).
So far, we have considered integers only for $q_\mu, q_\tau$. If they are
rational numbers, a discrete symmetry can force the zeroes to be exact.
For example, if $q_\mu= 1/3, q_\tau=-1/3$ then U(1)$_F$ breaks to a $Z_3$
gauge 
symmetry and there are {\em exact}\/ texture zeroes in the (1,1) and (2,2)
entries.

As an explicit viable example of corrected texture zeroes, we choose
$q_\mu=-2, q_\tau=2$:
\beq
M_R\sim M'' \left(\begin{array}{cc}
\epsilon^4 & 1 \\
1 & \epsilon^4 \\ \end{array} \right)
\eeq
from which we see that the desired structure is accurately
reproduced. $M''=10^{12}$ GeV gives the correct $\Delta m^2$ for this charge
assignment, and so could be the direct breaking scale of SU(2)$_R$,
generated for example in the several-stage intermediate breaking of a GUT\@.
In general $M''$ has an order of magnitude upper bound of the scale of
SU(2)$_R$ breaking
(or the breaking of a symmetry which contains SU(2)$_R$). 
If $M''$ has a non-renormalisable origin it can be orders of magnitude
lower than the SU(2)$_R$ scale. This can be the case if the tree-level mass is
forbidden by the vertical gauge symmetries, as in ref.~\cite{us} for example.
An example of the approximate form in Eq.(\ref{rhmaj}) was
found by appealing to
U(1)$_F$ gauge symmetry in ref.~\cite{neutpeeps}, where the authors
concentrate on the Giudice ansatz for Dirac masses. 

With $\epsilon \sim 0.1$, 
\beq
|q_\tau+q_\mu| \sim \log_{10} \frac{M''}{M}
\eeq
would predict the correct order of magnitude of $M$, given $M''$.
Using $M \sim 10^{12}$~GeV to explain the maximal mixing 
and $M'' = M_{GUT} =
10^{16}$~GeV, we observe that $|q_\tau+q_\mu|=4$ predicts the correct order
of magnitude for $M$ for this special case. 

\section{Conclusions}
We have shown that the maximal mixing of $\nu_\mu$, $\nu_\tau$ 
atmospheric neutrino anomaly can be naturally obtained 
in unified theories. 
Our basic assumption is that 
Dirac mass mixing is small, motivated by quark-lepton unification present in
unified theories. The maximal mixing occurs within the right handed
neutrino Majorana mass matrix.
The mechanism that motivates this ansatz is already familiar: a U(1)$_F$
family gauge symmetry
spontaneously broken at a high scale. 
The parameters of the oscillation measured by Super-Kamiokande
($\Delta m^2$
and large mixing angle) can be set successfully by a choice of family
dependent
quantum numbers.

The model is general in the sense that by only considering the two relevant
neutrinos, and not taking renormalisation effects into account, it is not
sensitive to 
the particular form of extended symmetry one wants to examine. Other neutrino
anomalies can be explained by oscillations
without conflict with this scheme
if their mixings are weak.
For example, to motivate the small-angle MSW solution to the solar neutrino
problem, one might include $\nu_e$ oscillations with small angles.
In particular, choosing the U(1)$_F$ charge of the right-handed electron
neutrino, $q_e=p$ (where p is an integer) and 
$q_\mu=1/3, q_\tau=-1/3$ would yield the exact form 
\beq
M_R = M'' \left( \begin{array}{ccc}
\lambda \epsilon^{2 |p|} & 0 & 0 \\
0 & 0 & 1 \\
0 & 1 & 0 \\
\end{array} \right),
\eeq
then the angle of the MSW-type oscillations would be naturally small, as they
originate from the small Dirac mixing. 
One conclusion of ref.~\cite{Bando:1998ns} will hold here also if we imposed
the Georgi-Jarlskog texture for down and charged lepton mass matrices,
i.e.\ that the
small $\sin^2 2 \theta_{e \mu}$
mixing required becomes compatible with the
size of $|V_{us}|$ because of suppression due to $\nu_\tau,
\nu_\mu$ maximal mixing.
However, the unknown dimensionless coupling $\lambda$ would have to be tuned
to make $\nu_e$ degenerate with $\nu_{\mu,\tau}$ in order to provide the low
$\Delta m^2$ values required by solar neutrino oscillations.
Another possibility would be to have an approximately massless $\nu_e$,
providing a candidate $\Delta m^2_{12}$ in the correct ballpark for the
reported LSND oscillations. Then perhaps a sterile neutrino could allow for
solar neutrino oscillations by mixing with the light $\nu_1$.
If there are other large mixings of the light neutrinos
(as could be implied by the vacuum oscillation or large angle MSW solutions to
the solar neutrino problem), the approximation of only considering two
neutrinos is incorrect. Much work~\cite{bimax} has recently been focussed upon
this possibility.

An obvious extension of the present scheme is to include $\nu_e$ and possibly
$\nu_s$ oscillations to explain the other data. 
An explicit treatment of the 
unified quark and
lepton U(1)$_F$ family charges would then be important to see whether the
solar and/or LSND data can be reproduced. 
The neutrinos responsible for the atmospheric neutrino anomaly are predicted
to be approximately degenerate, with masses of the order of 0.2 eV. 

\section*{Acknowledgements}
I would like to thank W. Buchmuller, H. Dreiner, C. K. Jung, S. F. King,
S. Sarkar and 
W. Scott for discussions about this work. 
This work was supported by PPARC and partly carried out at
Ringberg Euroconference `New Trends in Neutrino Physics 1998'.


\end{document}